\documentclass{tlp}

\usepackage{latexsym}

\usepackage{amsfonts}

\def\B{{\cal B}}
\def\H{{\cal H}}
\def\U{{\cal U}}

\def\UP{{\cal U}}

\newtheorem{defi}{Definition}
\newtheorem{theorem}{Theorem}
\newtheorem{ese}{Example}

\newcommand{\hide}[1]{}

\newcommand{\bb}{\tilde{b}}
\newcommand{\uu}{\tilde{u}}

\begin{document}

\title[Enhancing the Expressive Power of the U-Datalog
Language] {Enhancing the Expressive Power of the U-Datalog
Language}

\author[Elisa Bertino, Barbara Catania, and Roberta Gori]
{ELISA BERTINO \\ University of  Milano \\ Via
Comelico 39, 20135 Milano, Italy \\
 e-mail: bertino@dsi.unimi.it
 \and
BARBARA CATANIA \\ University of Genova \\ Via
Dodecaneso 35, 16146 Genova, Italy\\ e-mail: catania@disi.unige.it
\and ROBERTA GORI\\
 University of Pisa \\
 Corso
Italia, 40 \\
56125
Pisa, Italy \\
e-mail:
gori@di.unipi.it
}
\pagerange{\pageref{firstpage}--\pageref{lastpage}}
\volume{\textbf{1} (1):}
\jdate{January 2001}
\setcounter{page}{1}
\pubyear{2001}

\maketitle
\label{firstpage}


\begin{abstract}
U-Datalog has been developed with the aim of
providing a
set-oriented
logical
update language, guaranteeing update
parallelism in the
context of
a
Datalog-like language. In U-Datalog,
updates are expressed
by
introducing
constraints ($+p(X)$, to denote
insertion, and $-p(X)$,
to
denote
deletion) inside Datalog rules.
 A U-Datalog program can
be
interpreted as a CLP program. In this framework, a set of
updates
(constraints) is satisfiable if it does not represent
an
inconsistent theory, that is, it does not require the insertion
and the
deletion of the same fact. This approach resembles a very
simple form of
negation. However, on the other hand, U-Datalog
does not provide any
mechanism to explicitly deal with negative
information, resulting in a
language with limited expressive
power. In this paper, we provide a
semantics, based on
stratification, handling the use of negated atoms in
U-Datalog
programs and we show  which problems arise in defining
a
compositional
semantics.
\end{abstract}

\section{Introduction}

Deductive database technology represents an important step towards
the goal of
developing highly-declarative database programming
languages. Several
approaches for the inclusion of update
capabilities in deductive  languages
have been proposed. In
general, all those proposals are based on including
in rules, in
addition to usual atoms, special atoms denoting updates. In
most
of those proposals, an update execution consists of a query
component,
identifying the data to be modified, and  an update
component, performing
the actual modification on the selected
data. A way to classify  deductive
update languages is with
respect to the approach adopted for handling
possible
interferences between the query and update component of the
same
update execution. In particular, updates can be performed as soon
as
they are generated, as side-effect of the query evaluation,
thus, by
applying an {\em immediate} semantics. Languages based on
an immediate
semantics include $\cal LDL$ \cite{NT}, TL
\cite{AV91}, DL \cite{AV91}, DLP
\cite{MW88b}, Statelog
\cite{LL94}. The immediate semantics is in
contrast with the
{\em deferred update semantics}, by which updates are not
applied
as soon as they are generated during the query evaluation;
rather,
they  are executed only when the query evaluation is
completed.
Languages based on a deferred semantics include
U-Datalog
\cite{TKDE}, Update Calculus \cite{Chen95,Chen97}, and
ULTRA
\cite{WF97,WFF98}. Other languages, such as Transaction
Logic
\cite{BK94}, provide   both policies.

In this paper, we consider
U-Datalog, a language based on a
deferred semantics. Even if more
expressive and flexible
frameworks exist (see for example
\cite{BK94,WFF98}), the choice
of U-Datalog is motivated by the fact that
it represents an
immediate extension of Datalog to deal with updates. This
aspect
makes this language quite suitable for analyzing properties
related
to logical update languages \cite{BBC96}. In U-Datalog,
updates are
expressed by introducing constraints  inside Datalog
rules. For example,
$+p(a)$ states that in the new state $p(a)$
must be true where $-p(a)$
states that in the new state $p(a)$
must be false. Thus,
 U-Datalog
programs are formally modeled
as Constraint
Logic Programming
(CLP)
programs
\cite{JL87}.

In
CLP, any
answer to a given goal
(called a {\em
query}, in the database
context)
contains a set of
constraints,
constraining the resulting
solution. In
U-Datalog, each
solution contains
a
substitution for the query
variables
and a set of updates.
The execution
of a goal
is
based on a deferred
semantics.
In particular, given
a query,
all the solutions  are
generated in the so-called {\em
marking
phase},
using a CLP answering
mechanism. All the updates, contained
in the
various
solutions, are then
executed in the {\em update phase}, by
using
an
operational semantics.
The
set of all  updates  generated
during the
marking phase   forms
a
constraint theory which can be
inconsistent.  From
a logical point of
view, this
means that the update set
contains
constraints of the form
$+p(a)$, $-p(a)$,
requiring the insertion
and the
deletion
of the same
fact. The
U-Datalog computational
model rejects any
form of
conflict, both locally,
i.e., inside a single
solution, and
globally, from
different solutions. Thus, the set of updates to be executed
is always
consistent.

Besides the marking
and
update semantics phases, it
is often useful to devise an additional
semantics,
known as {\em
compositional semantics} \cite{TKDE}.
This
semantics, which is orthogonal
with respect to the one defined above,
characterizes  the semantics of
the
intensional
database
independently from the semantics of
the
extensional one and is based on the notion of
open
programs
\cite{BOGALE91}.
The compositional semantics is quite
important in the context
of deductive
databases since it provides  a
theoretical framework for
analyzing  the
properties of  intensional
databases.
Indeed, it
is always recursion free, even if it is not always
finite.
Therefore, when it is finite,
it also represents a useful
pre-compilation
 technique for
intensional databases.
However, since
this
semantics is usually
expensive to  compute, it is mainly used  for
analysis purposes.

Even if U-Datalog allows us to easily
specify
updates
and transactions, its
expressive power is limited since
no
negation
mechanism is provided, even if, due to update inconsistency,
some
limited form of
 negation on the extensional database is
provided.
This kind of negation
is
obviously
not sufficient to support a
large variety of user requests. In this paper we provide
an
operational
mechanism  handling negated atoms
in
U-Datalog programs, providing
a
marking phase and a
compositional
semantics. The proposed extension is
based
on the notion of
stratification,
first proposed for logic programming
and
 deductive
databases \cite{AV91,LL94,MW88b,NT}. This extension
is not, however, a straightforward extension of previously defined
stratification-based
semantics for two main reasons. First of all,
U-Datalog rules are
not range restricted
\cite{CGT} but are required to be
{\em safe through query
invocation},\footnote{See Section \ref{ssec:syntax}
for the
formal definition of this property.} resulting in  a
non-ground
semantics.  Note that,
even if
this is a
typical
assumption
in a real
context, most of the other
deductive
update
languages
require
range
restricted rules or interpret
free
variables as
{\em
generation of
new
values} \cite{AV91}.
A second
difference is that
 an
atom may fail not only
because an answer
substitution
cannot be found but
also because it
generates an inconsistent
set of updates.

In the
following,   we first
introduce U-Datalog  in
Section
\ref{sec:U-Datalog} and we extend it to
deal with
negation in Section
\ref{sec:neg}. Finally, in
Section
\ref{sec:concl} we
present some
conclusions and outline
future
work. Due to space limitations,
we assume the
reader to be aware of the basic
notions
of
(constraint) logic programming
\cite{JL87,Lloyd87} and
deductive
databases
\cite{CGT}. For additional
details on  U-Datalog,
see
\cite{TKDE}.

\section{U-Datalog}
\label{ssec:3}
\label{sec:U-Datalog}

\subsection{Syntax}
\label{ssec:syntax}

A
U-Datalog database
consists of: (i) an
extensional database
(or simply
database) $EDB$, that
is, a set of ground
atoms ({\em extensional atoms});
(ii) an
intensional
database
$IDB$ (or simply program), that is,
a set of rules of the form:\footnote{In the following, we
assume that
constants and multiple occurrences of the same variable
inside each
atom
are expressed by equality
constraints between
terms.}\\
\centerline{$H
\leftarrow
b_1, \ldots, b_k, u_1, \ldots, u_s,
B_1, \ldots, B_t
$}
\noindent where
$H$, $B_1, \ldots, B_t $ are atoms,
$b_1, \ldots, b_k$
are
equality
constraints, i.e. constraints of
type $X =
t$ (denoted by $\bb$), where $X$ is a
variable and
$t$ is a term, and
$u_1,
\ldots, u_s$
are
{\em update
constraints} (denoted by
$\uu$), also called {\em update atoms}.
An update
constraint is an extensional atom preceded by
the symbol
$+$, to denote an
insertion,
or by the symbol $-$, to denote a
deletion.

In the
following,
the set of extensional predicates is
denoted
by
$\Pi_{EDB}$, the
set of intensional predicates is denoted
by
$\Pi_{IDB}$, and
the Herbrand Universe
 is denoted
by $\cal H$
\cite{Lloyd87}. Moreover, a conjunction of equality
and update constraints
is simply called {\em
constraint}.\footnote{In the following, conjunction
between constraints is represented by using
`,' inside body rules, and by
using `$\wedge$' in other contexts.}
As
usual in deductive databases,
$\Pi_{EDB}$ and $\Pi_{IDB}$
are
disjoint.
Note that a U-Datalog program can
be seen as a CLP
program
where
constraints are represented by equalities
and update
atoms.

 A
U-Datalog
transaction is a goal.
In order to
guarantee a finite answer
to each goal and the
generation of a set
of
ground updates, we  assume
that rules are {\em
safe through
query
invocation}. This  means that, given
a U-Datalog
database $IDB$ and a
goal $G$, each
variable appearing in the
head or in
the update constraint
of a rule, used in
the evaluation of the
goal,
either appears in an atom contained in the body
of the same rule, or is
bound by a
constant
present in the goal. In this
case, $G$ is {\em
admissible}
for
$IDB$.

\begin{ese} \label{exe:syntax}
The
following
program is a
U-Datalog intensional
database:
\begin{tabbing}
$r1:rem\_man(X,Y) \leftarrow -dep\_A(Y),
emp\_man(X,Y) $\\
$r2:rem\_man(X,Y) \leftarrow
-dep\_A(Y),emp\_man(X,Z),rem\_man(Z,Y) $\\
$r3:ins\_man(X)\leftarrow
+dep\_A(X),rem\_man(X,Y)  $
\end{tabbing}
An atom $emp\_man(a,b)$ is true
if  `$b$' is a manager of `$a$'.
An atom  $rem\_man(a,b)$  is true if
`$b$'
is a (possibly indirect) manager of  `$a$'.
As a side effect,
it requires
the removal of `$b$' from department A.
An atom $ins\_man(a)$ is true if
`$a$' has at least one manager
and, as side effect, requires
the insertion
of `$a$' in department $A$.
At the same time, it requires the deletions of
all the (possibly indirect)
managers of `$a$' from department A.
\hfill
$\Diamond$
\end{ese}

\subsection{Semantics}
\label{ssec:semantics}

U-Datalog
constraints are interpreted over the Herbrand Universe
$\cal H$. In this
domain, equalities have the usual meaning:
$+p(\tilde{X})$ is interpreted
as the atom $p(\tilde{X})$ and
$-p(\tilde{X})$ is interpreted as the
negated atom $\neg
p(\tilde{X})$. If a constraint $\bb\wedge\uu$ is $\cal
H$-solvable,
i.e. if ${\cal H} \models \bb,\uu$,  there exists at least
one
substitution  that makes the constraint true. Thus, for no atom
both an
insertion and a deletion are simultaneously required. When
this is not
true, updates are said to be {\em inconsistent}.
The execution of ground
inconsistent update atoms
(e.g., $+p(a),-p(a)$) may lead to different
extensional
databases, with respect to the chosen execution order.

The
generation of inconsistent updates is avoided as follows:
 (i)
{\em
locally}: a solution containing an
inconsistent set of
updates (i.e.,
an unsolvable set of constraints) is not
included in the resulting set of
solutions
for the given
goal;
(ii) {\em globally}: if an
inconsistency
is
generated due to two consistent
solutions, the goal
aborts, no update
is
executed, and the database is left
in the state it had
before the
goal evaluation.

The
semantics of U-Datalog
programs is given
in two main steps.
In the first  step,
all solutions
for a given
goal are
determined   by
applying a CLP evaluation method ({\em
marking phase}, see
Section \ref{ssec:marking}).
Each solution
contains a set of bindings for
the
query
variables and a set of consistent update
atoms.
 In the second
step ({\em
update
phase}, see Section \ref{ssec:update}),
the
updates
gathered in the
various solutions are executed only if they
are
consistent.

Besides the marking and the update semantics, an
additional semantics is
sometimes introduced in the database context, which
is called {\em compositional
semantics} (see Section \ref{ssec:comp}).
Such
semantics characterizes the intensional database independently
from the
semantics of the extensional one. The compositional semantics is
quite
important in the context
of deductive
databases since it provides  a
theoretical framework for
analyzing  the
properties of  intensional
databases.

\subsubsection{Compositional
semantics}\label{ssec:compo}
\label{ssec:comp}

Since the extensional
database is the only
time-variant
component of a U-Datalog
database, for analysis purposes, it
is  useful to define the semantics
of a U-Datalog
intensional
database
independently from the current
extensional database.
Such semantics is called {\em compositional semantics} and
is always
represented by
 a  recursion free  set of rules.
Therefore, when it is
finite,
or when an equivalent finite set of rules
can be detected,
it
also represents a useful  pre-compilation
 technique for
intensional
databases.

The compositional  semantics
can be
defined assuming
the
intensional database to be an {\em open
program}
\cite{BOGALE91}, i.e.,
a program where
the knowledge regarding
some
predicates is assumed to
be
incomplete.
 Under this meaning, a
U-Datalog
intensional database can
be
seen as a program that is open with
respect to the
extensional
predicates.
The semantics of an open program is
a set of rules, whose
bodies  contain just open
predicates.
In order to define the compositional
semantics of a
U-Datalog
intensional
database, we introduce the following
set:\\
\centerline{$ID_{EDB}=\{
p(\tilde X) \leftarrow p(\tilde X)\; |\;
p\in
\Pi_{EDB}\}.$}
In
the previous expression, $\tilde{X}$ denotes a
list
of distinct variables.
Similarly to  \cite{TKDE},
we now  introduce  an
{\em
unfolding
operator}. Such operator,
given  programs
$P$ and
$Q$,
replaces an atom $p(\tilde{X})$, appearing in
the body of a
rule
in
$P$, with the body of a rule defining $p$
in
$Q$.

\begin{defi} Let
$P$ and $Q$ be  U-Datalog programs.
Then\footnote{In the
following, we
use the notation $\bigwedge_i c_i$
to
represent the conjunction
of
constraints $c_1 \wedge ... \wedge c_n$,
where
$n$ is clear from the
context.
The symbol $\equiv$
denotes
syntactic
equality.}

\begin{tabbing}
$Unf_{P}(Q)=\{$\=$p(\tilde X)
\leftarrow  \tilde{b}',\tilde{u}',
\tilde
H_{1},\ldots,\tilde H_{n} \;|
\exists$ a renamed rule\\
\>$p(\tilde X) \leftarrow \bb,\uu,p_{1}(\tilde
X_{1}),
\ldots,p_{n}(\tilde X_{n}) \in P$\\
\>  $\exists p_{i}(\tilde
Y_{i})\leftarrow \tilde
b_{i},\tilde
u_{i} , \tilde H_{i} \in
Q \
(i=1,...,n)$, which share no variables,\\
\>$ \tilde{b}'\equiv
\bigwedge_i
(\tilde b_{i}  \wedge
(\tilde X_{i}=\tilde Y_{i})) \wedge \bb$
\\
\>
$\tilde{u}'\equiv\bigwedge_{i}
\tilde u_{i}
\wedge \uu $\\
\>$\tilde{b}'\wedge \tilde{u}'$
is $\cal
H$-solvable$\}$\hspace*{7cm} $\Box$
\end{tabbing}
\end{defi}

The
compositional semantics of a U-Datalog
intensional
database  $IDB$ is
obtained by
repeatedly applying the
unfolding operator
until no new  rules
are generated.

\begin{defi}
 The
compositional
semantics
${\cal U}_{IDB}$ of $IDB$ with
respect to
$\Pi_{EDB}$ is defined as the
least
fixpoint
of
$
T_{IDB}^{c}(I)=Unf_{IDB}(I\cup ID_{EDB}). $
\hfill
$\Box$
\end{defi}

The
previous definition is based on the following
result, taken from \cite{TKDE}.

\begin{theorem}
 $T_{IDB}^{c}$ is
continuous. \hfill
$\Box$
\end{theorem}

\begin{theorem}
 For any
extensional
database $EDB$, for any
admissible
goal $G$, the evaluation of
$G$ in $IDB
\cup EDB$ generates the
same answer constraints than
the
evaluation of $G$ in
$\UP_{IDB} \cup EDB$.\hfill $\Box$
\end{theorem}
Note
that ${\cal
U}_{IDB}$ is always recursion free.
If
$IDB$ is a
recursive
program then ${\cal
U}_{IDB}$ in general  is not finite.
However, under
specific assumptions, it is equivalent to a finite set of
rules (see
Section \ref{ssec:8}).

\subsubsection{Marking
phase}
\label{ssec:marking}

The
answers to a U-Datalog query can be
computed in a
top-down or in
an
equivalent bottom-up style
\cite{JL87}.
Here,  we introduce only the
bottom-up semantics.
The
Constrained
Herbrand
Base $\cal B$ for a U-Datalog program is defined
as
the set of
{\em constrained}
atoms of the form $p(\tilde{X})
\leftarrow
b_1,...,b_k,u_1,...,u_n$
where $u_1,...,u_n$
are update atoms,
$b_1,...b_k$
are equality
constraints, $p \in \Pi_{EDB} \cup \Pi_{IDB}$,
and
$\tilde{X}$
is a tuple
of distinct variables. An interpretation is any
subset
of the
Constrained
Herbrand Base. Given a
U-Datalog database $DB =
IDB \cup EDB$,\footnote{Even if ground atoms contained in the extensional
database should
be represented as constrained atoms, we still write them as
ground
atoms to simplify the notation.}
operator
$T_{DB} : 2^{\cal
B}
\rightarrow 2^{\cal B}$ is defined
as
follows:\footnote{$2^{\cal B}$ is
the set
of all the  subsets of
the
Constrained Herbrand Base $\cal
B$.}

\begin{tabbing}
$T_{DB}(I)=\{
 p(\tilde X) $\=$\leftarrow \tilde
b',\tilde{u}' | \exists$
a renamed rule\\
\> $p(\tilde X)
\leftarrow
\bb,\uu,p_{1}(\tilde
Y_{1}),\ldots,p_{n}(\tilde
Y_{n})\in
DB$\\
\> $\exists
p_{i}(\tilde X_{i}) \leftarrow
\bb_{i},\uu_{i}\in
I \ (i=1,...,n),$ which share no
variables\\
\>
$\tilde
b'\equiv
\bigwedge_{i}(\bb_{i}\wedge (\tilde
X_{i}=\tilde
Y_{i}))\wedge \bb$\\
\>$\tilde{u}'\equiv\bigwedge_{i}
\uu_{i}\wedge \uu$
\\
\> $ \tilde{b}'\wedge
\tilde{u}'$
is $\H$-solvable
\hfill\}.\footnotemark
\end{tabbing}
\footnotetext{We assume that all  constraints generated by
a  fixpoint computation are projected onto the set of the head and
update atom variables. Moreover, we assume that a constrained atom
is inserted in the set being
constructed only if it is not redundant.}

\newpage

\begin{theorem}
Let $DB$ be a U-Datalog
database.
$T_{DB}$
is continuous
and admits  a unique least fixpoint ${\cal
FIX}_{DB}$
and
${\cal
FIX}_{DB} = T_{DB}{\uparrow
\omega}$.
Such fixpoint
represents
the bottom-up semantics of $DB$.\footnote{We recall
that $T_P
\uparrow 0
=
\emptyset$, $T_P \uparrow i =
T_P (T_P \uparrow
i-1)$,
$T_P{\uparrow
\omega}=
\bigcup_{i < \omega}
T_P \uparrow i$.} \hfill
$\Box$
\end{theorem}

Given
${\cal FIX}_{DB}$ and a goal
$G
\equiv
\leftarrow
\bb,\uu,p_1(\tilde{X}_1),...,p_n(\tilde{X}_n)$,
the
{\em
solutions} or {\em answer constraints}
for $G$
are
all
constraints
$\tilde{X}_1 =
\tilde{Y}_1,...,
\tilde{X}_n
=
\tilde{Y}_n,\tilde{b}',\tilde{u}'$
such
that
$
p_i(\tilde{Y}_i) \leftarrow \tilde{b}_i,\tilde{u}_i
\in
{\cal
FIX}_{DB}$
($i=1,...,n$), $\tilde{u}'\equiv
\uu \wedge
\tilde{u}_1\wedge...\wedge\tilde{u}_n$, $\tilde{b}'
\equiv
\bb\wedge\tilde{b}_1\wedge...\wedge\tilde
{b}_n$,  and
$\tilde{u}'\wedge\tilde{b}'\wedge
\tilde{X}_1
=
\tilde{Y}_1\wedge...\wedge
\tilde{X}_n
=\tilde{Y}_n$ is $\H$-solvable.
Let
$\tilde{b}'' \equiv \tilde{X}_1 =
\tilde{Y}_1\wedge...\wedge
\tilde{X}_n
=
\tilde{Y}_n\wedge\tilde{b}'$.
In this case, we
write
$G,DB
\stackrel{*}{\longmapsto}
\langle
\tilde{b}'',\tilde{u}'\tilde{b}''
\rangle$, where $\tilde{u}'\tilde{b}''$ denotes the result of the application of
the  equalities specified in $\tilde{b}''$ to $\tilde{u}'$. We assume that
$\tilde{b}''$ is restricted to the variables of $G$.
Note
that
$\tilde{u}'$
has to be
consistent.

\begin{ese}\label{EDB}
Consider
$EDB_i =
\{ emp\_man(b,b),\;emp\_man(b,c),\;
dep\_A(b), \;
dep\_A(c),\;dep\_B(b)\}$ and the
intensional
database of
Example
\ref{exe:syntax}. Transaction $T_1\equiv
\leftarrow
ins\_man(X)$
evaluated in
$EDB_i \cup IDB$ computes the consistent solution
$X =
b, Y =
c,
-dep\_A(Y),$ \linebreak $ +dep\_A(X)$. The additional solution $X = b,
Y
=
b,
-dep\_A(Y),+dep\_A(X)$ is not consistent and therefore
is
discarded by
the marking phase.\hfill
$\Diamond$
\end{ese}

\subsubsection{Update
phase}
\label{ssec:update}

 The {\em
update
phase} atomically executes the
updates collected by the
 marking
phase.
 Updates gathered
by
the
different
solutions for a given predicate are executed only
if
no
inconsistency arises. This guarantees that only
order
independent
executions are performed.
Formally, let $u =
\bigcup
\{\tilde{u}_j \mid
G,DB
\stackrel{*}{\longmapsto}
\langle
\tilde{b}_j,\tilde{u}_j \rangle \}$.
Let
$EDB_i$ be the current
database
state. If $u$ is a consistent and
ground set of
updates, the new database
$EDB_{i+1}$ is
computed  as
follows: $ EDB_{i+1}
= ( EDB_i \setminus
\{p(\tilde{t}) \mid
-p(\tilde{t})
\in u \}) \cup
\{p(\tilde{t}') \mid
+p(\tilde{t}') \in u\} $. In
this case,
we say that
$G$ {\em commits},
returning the tuple $\langle
\{\tilde{b}_j
\mid
G,DB
\stackrel{*}{\longmapsto}
\langle
\tilde{b}_j,\tilde{u}_j
\rangle\},
EDB_{i+1}, Commit \rangle$.
If $u$ is inconsistent or contains
at least
one non-ground
update atom,\footnote{Note that an unground set of updates can only be
generated by a non-admissible
goal.} we
let $EDB_{i+1}= EDB_i$ and
say that
$G$ {\em
aborts}. In this case, the
evaluation returns the tuple $\langle
\{\},
EDB_{i},
Abort
\rangle$.

\begin{ese}
Consider $EDB_i$ as in
Example \ref{EDB}  and
the intensional
database of
Example
\ref{exe:syntax}. The execution of
transaction $T_1
\equiv
\leftarrow ins\_man(X)$
generates the new
extensional
database
$EDB_{i+1}=
\{ emp\_man(b,b),\;emp\_man(b,c),$
$
dep\_A(b),\; dep\_B(b)\}.$
\hfill
$\Diamond$
\end{ese}

\section{Introducing
negation in
U-Datalog}\label{ssec:4}
\label{sec:neg}

Since solutions
containing
inconsistent updates are
not returned
by the marking phase,
 the
U-Datalog
semantics models some
kind of
negation.
This
form
of negation is
however very weak
with respect to the ability to
model
arbitrary
negation.
Indeed, it has been
proved that, with respect to
the returned
substitutions, U-Datalog is
equivalent to Datalog extended
with negation on
extensional predicates and
open with respect to a subset
of extensional
predicates \cite{potere}.
In the following, in
order
to
increase the expressive power of U-Datalog, we
introduce
{\em
negated
atoms} in the bodies of U-Datalog rules. The
resulting
language is
called
U-Datalog$^{\neg}$. Then, we assign a
semantics to such
language
when the
considered programs are stratified.
A
stratified U-Datalog$^{\neg}$
program is
defined
as
follows.

\begin{defi}\label{strati} A
U-Datalog$^{\neg}$
program $IDB$
is stratified if it
is possible to find a
sequence
$P_{1},\ldots,$
$P_{n}$, $P_i \subseteq IDB$ ($i=1,...,n$),
(also called stratification) such that the following
conditions
hold (in the following, we denote with  $Pred_i$  the set of
predicates defined in $P_i$):
\begin{enumerate}

\item
$P_{1},\ldots,P_{n}$ is a
partition of the
rules of $IDB$.
Each $P_i$ is called ``stratum''.

\item
For
each predicate
$q \in Pred_{j}$,  all the rules
defining $q$ in
$IDB$
 are in $P_{j}$.
 \item
If $q(u)\leftarrow \ldots,q'(v),\ldots \in IDB$, $q'\in Pred_{j}$,
then $q\in Pred_{k}$ with $j\leq k$. \item If $q(u)\leftarrow
\ldots,\neg q'(v),\ldots \in IDB$, $q'\in Pred_{j}$, then $q\in
Pred_{k}$ with $j< k$. \hfill $\Box$
\end{enumerate}
\end{defi}

The previous definition can be extended
to deal
with a U-Datalog
database
$DB = IDB \cup EDB$. In this case,
all
extensional facts belong to
the
first level.

In order to assign
a
semantics to stratified
U-Datalog$^{\neg}$ programs,   we
assume that
each
rule in the program is
safe through query invocation. Due
to
the
introduction of negation, the
notion of safety is extended by
requiring
that
each variable appearing in a
rule head, in a negated literal
contained
in a rule
body, or in an update
atom also appears in a positive
literal in
the rule body
or is bound by a
constant present in the goal.

The
main
differences between the bottom-up
semantics we are going to
present
and the bottom-up
semantics defined for
 Stratified
Datalog$^{\neg}$
programs \cite{CGT,CH85} are the
following.
Due to
the
condition of safety
through query
invocation, the semantics of
a
U-Datalog$^{\neg}$ program may
contain non-ground
constrained atoms
that,
however, will be made ground by
the goal.
Thus, negated atoms cannot
be
used, as usually done,
as
conditions to be satisfied by a solution.
Indeed,
 some variables inside the
generated solutions
may be made
ground
by the goal.  A solution to
this
problem is to
explicitly represent, during
the bottom-up computation,
the
solutions for
which a negated atom $\neg B$
is true.
 In this way, we
maintain all the
conditions that the
solutions
have to satisfy but the
check will be
executed only when a match with a
query goal is performed.
To represent
such
solutions, the underlying
constraint theory must be extended
to deal
with
inequality constraints of
type $X \neq a$, where $X$ is a variable
and
$a$
is a constant. For
example, if $X=a$ is the only solution for
$p(X)$,
then
$X \neq a$ is the
solution for $\neg p(X)$.\footnote{Note that, due to
the Closed
Word
Assumption
\cite{CGT},
this is
only a difference at the
presentation level
that
allows us to treat in an
homogeneous way positive and
negative
literals
during the bottom-up
computation.}

A second aspect is related to
the
semantics of $\neg B$
with respect to the updates
collected by
$B$.
$B$, in fact, can also fail due to the generation of inconsistent
updates.
Thus, all solutions containing  inconsistent updates  represent
solutions for
$\neg B$.
Solutions for
$\neg B$ in a
database
$IDB \cup EDB$
are therefore obtained by evaluating $B$ in
$IDB\cup
EDB$, and
complementing
not only   the computed constraints
but also the constraints
which ensure the consistency of the  updates generated
by evaluating $B$.

Finally,
we assume that
the derivation of $\neg
B$ does not
generate any update.
This assumption  is
motivated by the fact
that the
evaluation of $\neg B$
should be considered as a
test with respect
to the
bindings generated by
positive atoms.

In the following, we present the marking phase and the
compositional semantics
for Stratified U-Datalog$\neg$ programs. Note
that
no modification to the update
phase is
required.
Proofs of
the
presented results can be found in
\cite{TR}.

\subsection{The
marking phase semantics}\label{ssec:7}

As a natural
extension  of the
constraints
domain presented in \cite{TKDE},
the
Constrained Herbrand Base
for
U-Datalog$^{\neg}$ (denoted by
${\cal
B}^{\neg}$) consists of
constrained
literals of the form $L
\leftarrow  \tilde b,\tilde u$,
where $
\tilde b$ is a
conjunction of
equality and inequality
constraints, $ \tilde
u$
is a conjunction
of update atoms, and $L$ is a literal.
If $L$ is a
negated
atom,
$\tilde u$ is empty. In
the following, the set of all
conjunctions
of
equalities and inequalities
constraints, constructed on the
Herbrand Universe
$\cal H$, is denoted by
$\cal C$.

\begin{defi}
\label{def:tpneg}
Let $DB =IDB \cup EDB$ be a
U-Datalog$^{\neg}$
database.
The bottom-up
operator $T^{\neg}_{DB} :
2^{{\cal B}^{\neg}}
\rightarrow
2^{{\cal
B}^{\neg}} $ is defined
as
follows:
\begin{tabbing}
$T_{DB}^{\neg}(I)=\{ p(\tilde X) $\=$\leftarrow
\tilde b',\tilde
u' |\exists$ a renamed rule\\
\> $p(\tilde X)
\leftarrow
\bb,\uu,L_{1}(\tilde
Y_{1}),\ldots,L_{n}(\tilde Y_{n})\in
DB$\\
\> $\exists
L_{i}(\tilde X_{i}) \leftarrow \bb_{i},\uu_{i}\in
I \ (i=1,...,
n),$ which share no
variables\\
\> $\tilde
b'
\equiv\bigwedge_{i}(\bb_{i}\wedge (\tilde
X_{i}=\tilde Y_{i}))\wedge
\bb$ \\
\> $\tilde u' \equiv\bigwedge_{i}u_{i}\wedge
\uu$ \\
\> $\tilde
b'\wedge\tilde
u'$ is $\H$-solvable \hfill\}.
 \mbox{\hspace*{6.1cm}
\hfill
$\Box$}
\end{tabbing}
\end{defi}

Before introducing the
fixpoint
semantics, we
define an operator $Neg$ which
performs the negation
of a
constraint
belonging to ${\cal C}$.

\begin{defi}
Let $c = c_1 \wedge ... \wedge c_n$.
 $Neg:{\cal C}
\rightarrow
2^{\cal
C}$ is
defined as follows:
 \[Neg(c)=\left
\{
\begin{array}{ll}
\{  c_{1}', \ldots,   c_{m}'\} & c_{1}' \vee ...\vee c_{m}' \mbox{ is equivalent
  to } \neg c_{1} \vee ... \vee \neg c_{n} \\
                 & \mbox{$c_{i}'$ is $\cal H$-solvable }(i=1,...,m) \\
& \mbox{and }  \forall j, j=1,...,m, \\
&  \ c'_1 \vee ...\vee  c'_m \mbox{ is not equivalent to
}\\
& c'_1 \vee...\vee c'_{j-1}  \vee c'_{j+1} \vee ... \vee  c'_m, \\
\{ false\} & \mbox{otherwise \hspace*{6.1cm}
$\Box$} \\
\end{array}\right. \]
\end{defi}

For example, if $c \equiv X = 2 \wedge Y = 3$, then
$Neg(c) =
\{X \neq 2, Y \neq 3 \}$.
Operator $Neg$ is used to define an additional
operator
$Comp$,        which
takes a set  $S$ of
constrained positive
literals and returns
the set $H$ of
constrained
negative literals,
belonging   to the complement
of $S$. This
operator is
used to make
explicit the constraints for negative
literals at
the end
of the
computation of the positive literals of
each
stratum.

\begin{defi}\label{def:comp}
$Comp: 2^{B^{\neg}}\rightarrow
2^{B^{\neg}}$
is defined as
follows:
\begin{itemize}
\item $\neg
p(\tilde{X}) \leftarrow
\ \ \in
Comp(S)$ if there does not
exist any
$p(\tilde{Y}) \leftarrow \tilde
b,\tilde u  \in
S$.

\item $\neg
p(\tilde{X})  \leftarrow \tilde{b}' \in
Comp(S)$ iff $
p(\tilde{X})
\leftarrow
\tilde{b}_{1},\uu_{1},\ldots\ldots,
p(\tilde{X}) \leftarrow
\tilde{b}_{n},\uu_{n} $ are the only
(renamed apart) constrained
atoms for
$p$ in $S$,
$\overline{b'_{i}}\in
Neg((\tilde{b_{i}}\wedge
\tilde{bu})|_{\tilde{X}})$,\footnote{$c_{|X}$ denotes the
projection
of
constraint
$c$ onto the
variables in $X$ (thus, all the other
variables
are eliminated
by applying
a variable elimination
algorithm \cite{CH73}).}
\linebreak $\tilde{bu} \in
Sol(\tilde{u_{i}}\tilde{b_{i}}\footnote {See
note 9.})$
 ($i=1,...,n$),
$\tilde{b}'\equiv
\bigwedge_{i}\overline{b'_{i}}$,
and
$\tilde{b}'$ is
$\cal H$-solvable. \hfill $\Box$
\end{itemize}
\end{defi}

In the
previous definition, $Sol(\tilde u)$ is  the
  set of {\em minimal}
\footnote{Minimality is defined with
  respect to the order $\preceq$
defined as follows:
  $c\preceq c'$ if $\H\models c'\rightarrow c$}
constraints which implies that
$\tilde
u$ is a consistent set of
updates.
For example, $Sol(+p(a,Y),-p(X,Z),$ $-p(X,b),-p(b,c))=$
$\{X\not=a, Y\not=Z \wedge
Y\not =b\}$. Of course, if $\tilde{u}$ is an inconsistent set of update atoms,
no solution is generated and $Sol(\tilde{u}) = {false}$.
We also assume that all redundant constrained literals  contained in
$Comp(S)$ are removed.

In the previous definition,
operator $Neg$ is applied to
 computed constraints and to the constraints
which make satisfiable the
 updates generated by the corresponding positive
atom. Such solutions have been
restricted to the head  variables  since all
the other
variables are not needed to define the solutions for the
negated atom.

The fixpoint of
$T_{DB}^{\neg}$ is
computed as
follows.
First, the fixpoint of a given stratum is computed.
 Then, all the
facts that
have not been derived
are made
explicitly false. This
corresponds to
locally apply the CWA. Note that,
due
to stratification,
this approach is
correct since each predicate
is
completely defined in one
stratum.

\begin{defi} Let $DB = IDB
\cup EDB$  be a
stratified
U-Datalog$^\neg$ database.
Let
$(P_{i})_{(1 \leq i \leq n)}$ be
a
stratification for  $DB$.
The
bottom-up semantics of $DB$ is defined
as
${\cal
FIX}^{\neg}_{DB}=M_{n}$ where
the sequence $M_{1},\ldots,M_{n}$
is
computed
as follows: \\
\indent
$M_{1}=T_{P_{1}}^{\neg}\uparrow \omega
\cup
Comp(T_{P_{1}}^{
\neg}\uparrow
\omega)$\\
\indent
$M_{i+1}=T_{P_{i+1}
\cup
M_i}^{\neg}\uparrow
\omega\cup
Comp(T_{P_{i+1}\cup
M_i}^{\neg}\uparrow
\omega ),
1 <i \leq n.$
\hfill $\Box$
\end{defi}

\begin{theorem}
Let $DB$  be a
stratified
U-Datalog$^\neg$
database. ${\cal FIX}^{\neg}_{DB}$ can be
computed in a finite
number of
steps.\hfill $\Box$
\end{theorem}

Due to
some basic results presented in
\cite{CH85}, the bottom-up
semantics of a stratified
U-Datalog$^{\neg}$ database
is independent from the chosen
stratification.

The answers to a
given
U-Datalog$^{\neg}$ goal are
computed as described
for U-Datalog
programs in
Subsection \ref{ssec:marking},
by replacing
${\cal FIX}_{DB}$
with
$\cal{FIX}^{\neg}_{DB}$.

\begin{ese}\label{ex:1}
Consider the
extensional
database $EDB$ of Example \ref{EDB} and
the
U-Datalog$^{\neg}$ program $IDB$
obtained by adding the following rules to the ones presented in
Example
\ref{exe:syntax}:
\begin{tabbing}
$r4: change\_man(X)
\leftarrow -emp\_man(X,Y),dep\_B(X),dep\_A(Y) $\\
$r5:
change\_man(X)\leftarrow X=Y,+emp\_man(X,Y),dep\_B(X),
\neg
ins\_man(X)$
\end{tabbing}
An atom $change\_man(a)$ is now true if `$a$'
belongs to department B and if there
exists at least one employee in
department A. In this case, it removes all
managers of `$a$' belonging to
department A.
It is also true  if `$a$' belongs to department $B$
and it
has no manager.
In this case, the evaluation removes all managers of
`$a$'
belonging to department A
and makes `$a$' manager of itself.

A
possible stratification for $DB = IDB
\cup EDB$ is
the
following:
$P_{1}=EDB \cup \{r1,r2\},$ $P_{2}=
\{r3\}$,
$P_{3}=\{r4,r5\}.$
$\cal{FIX}^{\neg}_{DB}$ is computed
as
follows:\\
\begin{tabbing}
$T_{P_{1}}^{\neg}\uparrow \omega=
\{$\=$
emp\_man(X,Y) \leftarrow X=b,Y=b;
$\\
\> $emp\_man(X,Y) \leftarrow X=b,Y=c;$ \\
\> $dep\_A(X)\leftarrow X=b;$ $dep\_A(X)\leftarrow X=c;$ $dep\_B(X)\leftarrow
X=b;$ \\
\>$rem\_man(X,Y) \leftarrow X=b,Y=
b,- dep\_A(Y);$\\
\> $rem\_man(X,Y)
\leftarrow X=b,Y=c,-dep\_A(Y);
\hfill\}$\\
\\
$Comp(T_{P_{1}}^{\neg}\uparrow
\omega)=\{
\neg emp\_man(X,Y)
\leftarrow X \neq b ;$\\
\> $\neg
emp\_man(X,Y)
\leftarrow Y \neq b, Y \neq c;$\\
\> $\neg dep\_A(X) \leftarrow X \neq b, X\not = c$; $\neg
dep\_B(X) \leftarrow X \neq b;$\\
\> $\neg rem\_man(X,Y)
\leftarrow
X \neq b;$\\
\> $\neg
rem\_man(X,Y) \leftarrow Y \neq b,
Y \neq c;
\hfill
\}$ \\
\\
$M_{1}=
T_{P_{1}}^{\neg}\uparrow
\omega\cup
Comp(T_{P_{1}}^{\neg}
\uparrow \omega )$\\ \\
$T_{P_{2} \cup
M_1}^{\neg}\uparrow
\omega=\{ ins\_man(X)
\leftarrow X =b,Y=c,-dep\_A(Y),+dep\_A(X) \} \cup M_1$\\
$Comp
(T_{P_2 \cup
M_1}^{\neg}\uparrow \omega)= \{\neg ins\_man(X) \leftarrow
X \neq b
\}$\\
$M_{2}=
T_{P_{2}\cup M_1}^{\neg}\uparrow \omega \cup
Comp(T_{P_
{2}
\cup M_1}^{\neg}\uparrow
\omega)$ \\ \\
$T_{P_{3}\cup
M_2}^{\neg}\uparrow
\omega= \{
change\_man(X) \leftarrow X=b,Y=c,-emp\_man(X,Y);$\\
\>
$change\_man(X)
\leftarrow X=b,Y=b,-emp\_man(X,Y) \} \cup M_2$\\
$Comp(T_{P _{3}
\cup
M_2}^{
\neg}\uparrow
\omega)= \{\neg change\_man(X) \leftarrow X \neq
b \}$\\
{$ {\cal FIX}^{\neg}_P=M_3
=T_{P_{3}\cup
M_2}^{\neg}\uparrow
\omega
\cup Comp(T_{P_{3}\cup
M_2}^{\neg}\uparrow
\omega).
$}\\
\end{tabbing}
Note that rule r5
does not provide any additional answer for
predicate \linebreak
$change\_man$. Indeed, $\neg ins\_man$ generates the
constraint $X \neq b$
and $dep\_B$ generates the constraint $X=b$. Thus, the
whole constraint is
inconsistent.\hfill$  \Diamond$
\end{ese}

\subsection{Compositional
semantics}\label{ssec:8}

 The compositional semantics
for
U-Datalog
programs (Section \ref{ssec:compo})  was defined by using an
unfolding
operator
which replaces the  atom $p(\tilde{X})$ in the body of a
rule with
the
body of a
rule defining $p$.
 Problems arise
when
unfolding  negated
atoms.
Suppose we want
to unfold  $\neg
p(\tilde{X})$, then
the
disjunction
of the bodies of {\em all} the rules
that define
predicate
$p$ in the
compositional semantics has to be
negated.
Suppose
the
following rules represent  the compositional
semantics
of a predicate
$p$:\\
$\begin{array}{c}
p(\tilde{X})
\leftarrow
\tilde{b}_{1},\tilde{u}_{1},L_{1},\\
\vdots
\\
p(\tilde{X})
\leftarrow
\tilde{b}_{n},\tilde{u}_{n},L_{n}.
\end{array}$

 Since
$p(\tilde{X})$ is true
(due to the CWA) if and only
if
$
\bb_{1},\uu_{1},L_{1}\vee \ldots\vee
\bb_{n},\uu_{n},L_{n}$ is
true,
$\neg
p(\tilde{X})$ has to be unfolded with
the negation
of
$\bb_{1},\uu_{1},L_{1}\vee
\ldots\vee
\bb_{n},\uu_{n},L_{n}$ (since
$\neg
p(\tilde{X}) \leftrightarrow
\neg
(b_{1},u_{1},L_{1}\vee
\ldots\vee
b_{n},u_{n},L_{n})$).\footnote{The way we  unfold a negated
atom
$\neg p(X)$ corresponds to
the syntactic transformation performed
in
the
Clark's completion approach \cite{Clark}.
 However, while Clark's
completion is
 used as a logical theory, our  resulting unfolded
program is
evaluated by using a
bottom-up stratified semantics. Therefore, we can
prove
 that $P$ and its
unfolded version are equivalent w.r.t
answer
constraints (see Theorem
\ref{pippo}).
Moreover,  since we
deal
with
stratified programs, no $L_{i}$
in the formula $f\equiv
p(\tilde{X})
\leftrightarrow
(b_{1},u_{1},L_{1}\vee \ldots\vee
b_{n},u_{n},L_{n})$ can be
equal
to $\neg p(\tilde{X})$,
since no cycle
through negation arises in predicate
definition. Thus,
$f$ is always
{\em
consistent} \cite{Clark}.
}

A problem arises when
there
exists
an
infinite set of rules defining $p(\tilde{X})$
in
the
compositional
semantics. In this case, the unfolding
operator cannot
be
applied since it is not effective.
In order to  solve
this problem,
a
weaker notion of compositionality can
be introduced, based on the restriction of the set of extensional
databases with
respect to which
the
intensional database can be composed. The
additional
information
available
on the considered extensional database has
to
guarantee that  the
result of
the unfolding operator, which unfolds all
the
positive literals
in the rule
bodies, is finite.
Gabbrielli et al. in
\cite{GabbrielliGM93}
showed that,
when the Herbrand Universe
$\cal H$ is
finite, it is possible to
compute a
T-stable semantics of a
logic
program $IDB$, which is finite and
gives
the
same
answer constraints of
$IDB$ when composed
with
any
extensional database defined
on $\cal H$. Intuitively, the
T-stable
semantics
iterates the unfolding
operator as many times as the new
unfolded
rules may
give  different results
on the finite domain $\cal
H$.\footnote{Note that
the finite domain
assumption can be
guaranteed only  by
executing updates
which do not insert
new values inside
the database.}

Under these hypothesis, the
compositional
semantics for a stratified
U-Datalog$^\neg$
program
corresponds to an unfolding
semantics computed in
two
steps:
\begin{enumerate} \item In the first step,
all positive literals
in
the rule
bodies are unfolded, by computing the
T-stable semantics
according
to the
algorithm given in
\cite{GabbrielliGM93} and the unfolding
operator
presented in
Section
\ref{ssec:compo}. At this stage, negative
literals are
left
unchanged.
\item In the second step, negative literals
are unfolded.
Due to
the finite
domain assumption and results
presented
in
\cite{GabbrielliGM93}, the set of
rules required to unfold
negative  literals
is
finite.
\end{enumerate}
 The result is a recursion
free program written in an {\em extended}
U-Datalog$^{\neg}$ language
which
characterizes the semantics of the intensional
database w.r.t.$\;$ the
extensional one.

\subsubsection{Unfolding of positive
literals}

In
order to
compute the compositional semantics of a program
$IDB$, we
first
unfold
positive literals by dealing with negative literals
as if
they
were
extensional predicates. This means that negative literal
are
not
unfolded.
To this purpose, the
techniques
presented
in
\cite{GabbrielliGM93} are applied to obtain
a
finite set of
rules,
denoted by  ${\cal U}^{pos}_{IDB}$. The basic idea
of
the
T-stable
semantics is illustrated by the
following
example.

\begin{ese}\label{ex:3}
Consider rules r1 and r2  presented in
Example
\ref{exe:syntax}
and suppose that $\H = \{a,b\}$. After
two
iterations of
the unfolding
operator presented in Section
\ref{ssec:compo},
we obtain the
following rules
(call
them
$T$):
\begin{tabbing}
$rem\_man(X,Y) \leftarrow
-dep\_A(Y),
emp\_man(X,Y)$\\
$rem\_man(X,Y)
\leftarrow  -dep\_A(Y),emp\_man(X,Z),
emp\_man(Z,Y).$ \end{tabbing}
At
the third iteration of
the unfolding
operator we also obtain the rule

\begin{tabbing}
$rem\_man(X,Y)
\leftarrow
$\=$-dep\_A(Y),emp\_man(X,Z),$\\
\>$emp\_man(Z,W),emp\_man(W,Y)$
\end{tabbing}

\noindent which cannot infer
different results on any
database defined on
two elements, since
$rem\_man(X,Y)$ computes the transitive closure of
relation
$emp\_man$. Thus,
 $T$ corresponds
to the  T-stable semantics
of
the previous
rules.
Now suppose that
$\H = \{a,b,c\}$. The T-stable
semantics
$\UP^{pos}_{IDB}$ is
computed
as
follows (in the following,
$\UP^{pos}({P_j})$  denotes the set of rules contained in
stratum
$j$
of
$\UP^{pos}_{IDB}$):
\begin{tabbing}
$\UP^{pos}_{IDB}=$\=$\UP^{pos}(P_1):$\=$rem\_man(X,Y)
\leftarrow
-dep\_A(Y),emp\_man(X,Y)$\\ \>\>$rem\_man(X,Y)
\leftarrow
-dep\_A(Y),emp\_man(X,Z), emp\_man(Z,Y)$\\ \>\>
$rem\_man(X,Y)
\leftarrow $\=$-dep\_A(Y),emp\_man(X,Z), $\\ \>\>
\>
$emp\_man(Z,W),emp\_man(W,Y)$\\ \> $\UP^{pos}(P_2):ins\_man(X)
\leftarrow
+dep\_A(X),-dep\_A(Y), emp\_man(X,Y)$\\ \>\>
$ins\_man(X)
\leftarrow +dep\_A(X),-dep\_A(Y),emp\_man(X,Z),$\\
\>
\>\>$emp\_man(Z,Y)$\\ \> \>$ins\_man(X)
\leftarrow
+dep\_A(X),-dep\_A(Y),emp\_man(X,Z),$\\
\>\>
\>$emp\_man(Z,W),emp\_man(W,Y)$\\ \>$\UP^{pos}(P_3): change\_man(X)
\leftarrow
-emp\_man(X,Y),dep\_B(X),dep\_A(Y) $\\ \>
\>$change\_man(X) \leftarrow
X=Y,+emp\_man(X,Y),dep\_B(X),$\\
\>\>\>$\neg ins\_man(X).$  \hspace*{4.2cm}
$\Diamond$
\end{tabbing}
 \end{ese}

By
results presented
in
\cite{GabbrielliGM93} and \cite{Ma93},
we can state the
following
results.\\

\begin{theorem}
Let $\H$ be the fixed and finite Herbrand
Universe. Let $\UP^{pos}_{IDB}$ the T-stable semantics computed as described in
\cite{GabbrielliGM93}, depending on the cardinality of $\H$.
For any
extensional database $EDB$,
$\UP^{pos}_{IDB}\cup EDB $ is
equivalent to
$IDB \cup EDB $. Moreover,
    $\UP^{pos}_{IDB}$
admits
the
same stratification of $IDB$ and
preserves
goal
admissibility.\hfill
$\Box$
\end{theorem}

\subsubsection{Unfolding
of
negative
literals}

After constructing ${\cal U}^{pos}_{IDB}$,
negative
literals
have to be unfolded.
In order to unfold a negated
literal
$\neg
p(\tilde{X})$,  the
disjunction of the bodies of {\em all}
the
rules
defining predicate $p$ in
${\cal U}^{pos}_{IDB}$ has to
be
negated.
This
approach should be applied stratum by stratum,
generating
in
a finite
number of steps a set of rules not containing
negative
literals.
Note that,   due to stratification conditions,
the
unfolding of
$\neg p$  is required only in rules belonging to levels
higher
than the
level where $p$ is defined.   The resulting  set
of rules
corresponds to
the compositional semantics of
$IDB$.
However,
unfortunately, the negated
disjunction of the bodies
defining $p$
is, in general, a first order formula, which cannot
 be
represented
in U-Datalog$^{\neg}$, as the
following
example
shows.

\begin{ese}\label{ex:pluto}
Consider  the intensional predicate $p$ defined by the rule
$r:p(X) \leftarrow X=a, f(X,Y),$ \linebreak$ q(X,Y)$, where $f$ and $q$ are
extensional predicates. The previous rule is logically equivalent
to the following first order formula: $p(X)\leftarrow X=a \wedge
\exists Y \ (f(X,Y) \wedge q(X,Y))$. By assuming that $r$ is the
only rule defining $p$, by CWA, we obtain that $\neg
p(X)\leftrightarrow \neg(X=a \wedge \exists Y \ (f(X,Y) \wedge
q(X,Y)))$. But $\neg(X=a \wedge \exists Y \ (f(X,Y) \wedge
q(X,Y)))$ is logically equivalent to  $(X\not = a) \vee (\forall Y
\ (\neg f(X,Y) \vee \neg q(X,Y)))$, which can always be
transformed in prenex disjunctive normal form
\cite{Maher}  obtaining
$\forall Y(X\not =a \vee \neg f(X,Y)\vee \neg q(X,Y))$. \hfill
$\Diamond$
\end{ese}

From the previous example it follows that, in order to unfold negative literals,
the
U-Datalog$^{\neg}$ syntax has to be
extended
to deal with first order formulas.
As shown in the example,  the variables
which
become
 quantified after negation, correspond
to {\em local
variables} of the
original rule, i.e., body
variables not appearing in the rule
head.
After this extension,
the syntax of a
U-Datalog$^{\neg}$
rule,
hereafter
called {\em extended U-Datalog$^\neg$}
rule,  becomes
the
following:\\
\centerline{$H
\leftarrow
\bb,\uu,\tilde{L} \diamond
\tilde{Q}(b_{1}, \tilde{H}_1 \vee ... \vee
b_{n},   \tilde{H}_n  )$}
 where $H$ is an
atom, $\uu$ is an update constraint, $\bb$ is a conjunction of
equality constraints, $\tilde{L}$ is a conjunction of positive
literals, $\tilde{Q}(b_{1}, \tilde{H}_1 \vee ... \vee
b_{n},\tilde{H}_n)$ is a first order formula in prenex disjunctive
normal form, where $\tilde{Q}$ is a sequence of quantified
variables, not appearing in $H$ or in $\tilde{L}$, each $b_{i}$ is
a conjunction of equality and inequality constraints, each
$\tilde{H}_i$ is a conjunction of literals.\footnote{Note that the
proposed extensions are performed at the body rule level. No
change to the Herbrand Base is performed.} Intuitively,
$\tilde{b},\tilde{u},\tilde{L}$ is generated by the unfolding of
positive literals whereas the first order formula
$\tilde{Q}(b_{1},\tilde{H}_1 \vee ... \vee b_{n},\tilde{H}_n)$ is
generated by the unfolding of negative literals.
Of course, we still assume that rules are stratified.

An extended U-Datalog rule
body  is true in a given interpretation
if there exist some bindings for the
positive literals and some bindings for the  free
variables of the quantified  formula which make the rule body true in the
given interpretation.
Formally,
the truth of an extended U-Datalog rule body can be defined
as follows.

\begin{defi}
Let $R \equiv
\bb,\uu,L_{1}(Y_{1}),
\ldots,L_{m}(Y_{m})\diamond F$ where
$F \equiv Q(b_{1},
\tilde{H}_1
\vee ... \vee b_{n},\tilde{H}_n)$.
Let $X_1.,,,.X_n$ be the free variables of $F$.
Let $I \subseteq \B^{\neg}$.
$R$ is true in $I$
with
answer
constraint $\overline{{b}},\overline{{u}}$ if
there
exist
$L_{i}(\tilde{U}_{i}) \leftarrow
b_{i},u_{i}\in I$ $(i= 1,\ldots,m)$, and
$c\equiv (X_{1}=t_{1}
\wedge \ldots\wedge X_{n}=t_{n})$ ($t_{j}\in \H, \ j=1,...,n$),
such that
$ \overline{b}\equiv   \bigwedge_i b_{i}\wedge \bb\wedge
c
\;\bigwedge_{i} (\tilde{Y}_{i}=\tilde{U}_{i})$,
$\overline{u}
\equiv \uu\
\wedge
(u_{1}\wedge   ...\wedge u_{m})$,
$\overline{{b}}\wedge\overline{{u}}$  is ${\mathcal H}$-solvable,
and $I\models
\overline{{b}}\wedge
F$.\footnote{$I\models
\overline{{b}}\wedge
F$ means that for any
assignment of values to quantified variables, satisfying
$\overline{b}$, it is
possible to find some constrained literals in $I$ unifying with those in $F$, such that
the resulting constraint is $\H$-solvable.} \hfill $\Box$
\end{defi}

The  bottom-up operator of
an
extended
U-Datalog$^{\neg}$ program can be now defined as
follows (in the following,  $body(r)$ denotes the body of a rule
$r$).

\begin{defi} Let $DB =IDB \cup EDB$ be
an
extended
U-Datalog$^\neg$ database.
The bottom-up operator $T^{e}_{DB}
:
2^{{\cal
B}^{\neg}} \rightarrow 2^{{\cal
B}^{\neg}} $ is defined
as
follows:
\begin{tabbing}
$T_{DB}^{e}(I)=\{$\=$p(\tilde X) \leftarrow
\overline{b},\overline{u} | \exists$
 a renamed rule\\
\> $r  : p(\tilde
X) \leftarrow \;
\bb,\uu,\tilde{L}\diamond \tilde{Q}(b_{1}, \tilde{H}_1 \vee ...
\vee
b_{n},\tilde{H}_n) \in DB $\\
\> $\overline{b},\overline{u}$ is an
answer constraint for  $body(r)$ in
$I$\}
\hspace*{3.5cm}
$\Box$
\end{tabbing}
\end{defi}

\begin{theorem}
Let $\H$ be a
finite
domain.
$T_{DB}^{{e}}$ is a
continuous
operator.\hfill
$\Box$
\end{theorem}

Given an extensional
database $EDB$ and
an
extended
U-Datalog$^{\neg}$ program $IDB$, the
semantics of $DB =IDB
\cup
EDB$ is obtained as
the least fixpoint of
$T_{DB}^{e}$, denoted by
${\cal
FIX}^e_{DB}$.
Due
to
the
presence of a first order formula   in rule bodies,
a new safeness
through
query invocation  property
has to be
 stated.
\begin{defi}
Let $P$
be an
extended Datalog$^{\neg}$ program.
$P$ is  {\em safe
through
query
invocation} if
each non-quantified variable, appearing  in a rule head,
in
an
update atom, or in a negated atom,
also
appears in
a
positive literal of the rule
body or is bound by a constant
present in
the goal.\hfill
$\Box$
\end{defi}

The unfolding   operator we are going to define
works stratum by stratum. First, positive
literals in the rule bodies
of each stratum are
unfolded by using  operator $\U^{pos}_{IDB}$.
Then, the
negative literals contained in the $i$-th stratum of $\U^{pos}_{IDB}$
are unfolded by
using an  operator $\U^{neg}_{IDB}$ and the rules resulting from
the
completed unfolding of   strata $1,...,i-1$.
The result is an extended
U-Datalog$^{\neg}$ program equivalent to $IDB$
 but not containing positive
or negative intensional literals.

Before presenting the unfolding
operator,
  we define the operator
$\U^{neg}_{IDB}$ which
unfolds the
negative literals of a   U-Datalog$^{\neg}$  program
  $IDB$, by using the
rules of an extended U-Datalog$^{\neg}$  program $U$.
 In order to
define
function $\U^{neg}_{IDB}$, we need an operator, $Neg_c$, which takes
the disjunction of a set
 of (extended) U-Datalog$^{\neg}$ rule bodies
defining a predicate
 $p$,
 performs its logical negation and  returns  the
resulting first order
 formula in prenex disjunctive normal form. Such formula is then used to
 construct the unfolded rule. Information on  the
head variables of each rule defining $p$ is useful to
 understand which are the local variables, that
is, the
 variables which have to be  quantified.
 In performing negation,
also
   constraints which make the
 update atoms
satisfiable have to be
considered, similarly to what
has been done for operator
$Comp$
(see Definition \ref{def:comp}).

\begin{defi}
Let
$P^e = \{IDB | IDB $ is an
extended
U-Datalog$^{\neg}$ program\}.
Let
$IDB$ be a stratified
U-Datalog$^{\neg}$
intensional database. The
unfolding
operator
$\U^{neg}_{IDB}:  P^e
\rightarrow P^e$ is defined
as
follows:
\begin{small}
\begin{tabbing}
$\U^{neg}_{IDB}$\=$(U)=\{p$\=$(\tilde
X) \leftarrow
 \bb,\uu, A_{1},\ldots,A_{n},\diamond F|\exists$ a
renamed rule\\
\>$p(\tilde X) \leftarrow \bb,\uu,A_{1},\ldots,A_{n},
\neg
p_{1}(\tilde
Z_{1}),\ldots,\neg
p_{m}(\tilde Z_{m}) \in IDB$\\
\>for
all $p_{i}$
($i=1,...,m$),
consider the body of all the rules $r^i_{j}$ ($j=1,...,l_i$) \\
\> \ \ defining
$p_{i}$ in
$U$\\
\>$r^i_1: p_{i}(\tilde
V_{i,1})\leftarrow\bb_{i,1},\uu_{i,1},\tilde{L}_{i,1}
\diamond F_{i,1}
\in U$\\
\>$\vdots$\\
\>$r^i_{l_i}:
p_{i}(\tilde
V_{i,{l_i}})\leftarrow
\bb_{i,{l_i}},\uu_{i,{l_i}},\tilde{L}_{i,{l_i}}
\diamond F_{i,{l_i}}  \in U$\\
\>$ F\footnotemark\equiv
\tilde{Q} (c_{1},\tilde{L_{1}}\vee \ldots \vee c_{e},\tilde{L_{e}})\in$ \\
\> \>
$Neg_{c}
($\=$(\tilde{Z_{1}}=\tilde{V}_{1,1}\wedge body(r^1_1)
\vee ... \vee \tilde{Z_{1}}=\tilde{V}_{1,l_{1}}\wedge body(r^1_{l_1}))
\wedge ... \wedge$ \\
\>\>\> $(\tilde{Z_{m}}=\tilde{V}_{m,1}\wedge body(r^m_1)
\vee ... \vee \tilde{Z_{m}}=\tilde{V}_{m,l_{m}}\wedge body(r^m_{l_m})))$\\
\> $\bb \wedge \tilde{u} \wedge \tilde{Q}(c_{1}\vee \ldots \vee c_{e})$
is ${\cal H}$-solvable $  \}$
\hspace*{5.8cm} $\Box$
\end{tabbing}
\end{small}
\end{defi}
\footnotetext{Note that no updates are
generated. }

By using operator $\U^{neg}_{IDB}$, the
compositional semantics
is defined as follows.

\begin{defi}\label{compo}
Let $IDB$ be a
stratified
U-Datalog$^{\neg}$ program.
Suppose that $IDB$,
and
therefore
$\UP^{pos}_{IDB}$, admits a stratification with $k$
strata
$P_1,...,P_k$.
Let
$\UP^{pos}({P_j})$  be the set of rules contained in
stratum
$j$
of
$\UP^{pos}_{IDB}$. The compositional semantics $\UP_{IDB}$
is
defined
as
follows (see Subsection \ref{ssec:comp} for the definition of $ID_{EDB}$):

\[\begin{array}{l}
\overline{P}_{1}=\U^{neg}_{\UP^{pos}{(P_1)}}(ID_{EDB})
\\
\overline{P}_{2}=\U^{neg}_{\UP^{pos}({P_2})}(ID_{EDB}\cup
\overline{P}_{1})\\
\;\;\;\vdots\\
\overline{P}_{i}=U^{n
eg}_{\UP^{pos}(P_{i})}(ID_{EDB}\cup
\overline{P}_{i-1})\\
\UP_{IDB}=\bigcup_{1\leq
i\leq k}
\overline{P}_{i}.
\mbox{\hspace*{9cm} \hfill
$\Box$} \end{array}\] \end{defi}

\begin{theorem}\label{pippo}
$\UP_{IDB}$ is safe
through
query
invocation and for each admissible goal $G$ and for each extensional database
EDB,
the answer
constraints
for $G$ in ${\cal
FIX}^{\neg}_{IDB} \cup EDB$ are the same than the ones
in
${\cal
FIX}^e_{\UP_{IDB}}\cup EDB$.\hfill $\Box$
\end{theorem}

\begin{ese}
Consider the
program
$IDB$ and its stratification, as  presented in Example
\ref{ex:1}, and
its
positive
unfolding
$\UP^{pos}_{IDB}$, as presented
in  Example
\ref{ex:3}.
The
compositional
semantics of $IDB$ is constructed
as follows:

\begin{tabbing}
$\overline{P}_{1}=$\=$ P_{1}$\hspace*{0.5cm}
$\overline{P}_{2}=$\=$P_{2}$ \\
$\overline{P}_{3}=$ \>  $change\_man(X) \leftarrow$
\= $
-emp\_man(X,Y),dep\_B(X),dep\_A(Y) $\\
\> $change\_man(X) \leftarrow X =
Y,+emp\_man(X,Y),dep\_B(X)\diamond$\\
\>\>$\forall Z (X=Z
\vee\neg
emp\_man(X,Z)) $\\
$\UP_{IDB}=\overline{P}_{1}\cup
\overline{P}_{2}\cup
\overline{P}_{3}.$
\end{tabbing}
The second rule  for predicate
$change\_man$
derives from the unfolding of  \linebreak predicate $\neg ins\_man(X)$.
The  formula \small{$\mathbf  \forall Z (X=Z
\vee\neg
emp\_man(X,Z)) $} \normalsize is the simplified  result of
the
application of the $Neg_{c}$ operator
 to the disjunction of the
bodies of the rules
defining  $ins\_man$.
\hfill
$\Diamond$
\end{ese}

It is
important to remark that
the
compositional semantics has not to be
considered as an alternative semantics
w.r.t. the marking phase
semantics.
Indeed, the computation of the compositional semantics can  be
quite
expensive.
However, since the compositional semantics is recursion
free and
has to be computed just once (unless the Herbrand domain
changes)
it can be meaningfully used in some cases  as a precompilation
technique.

\section{Concluding remarks}
\label{sec:concl}

In this
paper we
have
introduced negation
inside U-Datalog rules
and proposed
a
stratification-based approach to
assign a
semantics to such programs.
We
have also introduced a weaker concept
of
compositionality and presented
a finite and
effectively
computable
compositional semantics
for
U-Datalog$^\neg$
programs.
By results presented in \cite{potere}, it is
quite
immediate to
prove that, with respect to the returned
answers,
U-Datalog$^{\neg}$ is
equivalent to Stratified Datalog$^{\neg}$,
open
with respect to a subset of extensional
predicates  \cite{TR}.
The
presented results  can
be
extended to deal with
other Datalog-like update
languages safe through
query
invocation. Future
work includes the
introduction of negation in
other
U-Datalog
extensions
\cite{BCRV98a,BCRV98b} and the definition of
static
analysis
techniques for
U-Datalog$^\neg$, similarly to
those
proposed for
U-Datalog
\cite{BBC96}.

\label{lastpage}

\end{document}